\documentclass[twocolumn,aps,unsortedaddress,superscriptaddress,prl,floatfix,showpacs]{revtex4}
\usepackage{epsfig}

\begin{document}


\title{Measurement of Angular Distributions of Drell-Yan Dimuons
in $p + p$ Interactions at 800 GeV/c}

\affiliation{Abilene Christian University, Abilene, TX 79699}
\affiliation{Physics Division, Argonne National Laboratory, Argonne, IL 60439}
\affiliation{Fermi National Accelerator Laboratory, Batavia, IL 60510}
\affiliation{Georgia State University, Atlanta, GA 30303}
\affiliation{Hampton University, Hampton, VA 23187}
\affiliation{Illinois Institute of Technology, Chicago, IL  60616}
\affiliation{University of Illinois at Urbana-Champaign, Urbana, IL 61801}
\affiliation{Los Alamos National Laboratory, Los Alamos, NM 87545}
\affiliation{University of New Mexico, Albuquerque, NM 87131}
\affiliation{New Mexico State University, Las Cruces, NM 88003}
\affiliation{Oak Ridge National Laboratory, Oak Ridge, TN 37831}
\affiliation{Texas A\&M University, College Station, TX 77843}
\affiliation{Valparaiso University, Valparaiso, IN 46383}

\author{L.Y.~Zhu}
\affiliation{Hampton University, Hampton, VA 23187}
\affiliation{University of Illinois at Urbana-Champaign, Urbana, IL 61801}

\author{J.C.~Peng}
\affiliation{University of Illinois at Urbana-Champaign, Urbana, IL 61801}
\affiliation{Los Alamos National Laboratory, Los Alamos, NM 87545}

\author{P.E.~Reimer}
\affiliation{Physics Division, Argonne National Laboratory, Argonne, IL 60439}
\affiliation{Los Alamos National Laboratory, Los Alamos, NM 87545}

\author{T.C.~Awes}
\affiliation{Oak Ridge National Laboratory, Oak Ridge, TN 37831}

\author{M.L.~Brooks}
\affiliation{Los Alamos National Laboratory, Los Alamos, NM 87545}

\author{C.N.~Brown}
\affiliation{Fermi National Accelerator Laboratory, Batavia, IL 60510}

\author{J.D.~Bush}
\affiliation{Abilene Christian University, Abilene, TX 79699}

\author{T.A.~Carey}
\affiliation{Los Alamos National Laboratory, Los Alamos, NM 87545}

\author{T.H.~Chang}
\affiliation{New Mexico State University, Las Cruces, NM 88003}

\author{W.E.~Cooper}
\affiliation{Fermi National Accelerator Laboratory, Batavia, IL 60510}

\author{C.A.~Gagliardi}
\affiliation{Texas A\&M University, College Station, TX 77843}

\author{G.T.~Garvey}
\affiliation{Los Alamos National Laboratory, Los Alamos, NM 87545}

\author{D.F.~Geesaman}
\affiliation{Physics Division, Argonne National Laboratory, Argonne, IL 60439}

\author{E.A.~Hawker}
\affiliation{Texas A\&M University, College Station, TX 77843}

\author{X.C.~He}
\affiliation{Georgia State University, Atlanta, GA 30303}

\author{L.D.~Isenhower}
\affiliation{Abilene Christian University, Abilene, TX 79699}

\author{D.M.~Kaplan}
\affiliation{Illinois Institute of Technology, Chicago, IL  60616}

\author{S.B.~Kaufman}
\affiliation{Physics Division, Argonne National Laboratory, Argonne, IL 60439}

\author{S.A.~Klinksiek}
\affiliation{University of New Mexico, Albuquerque, NM 87131}

\author{D.D.~Koetke}
\affiliation{Valparaiso University, Valparaiso, IN 46383}

\author{D.M.~Lee}
\affiliation{Los Alamos National Laboratory, Los Alamos, NM 87545}

\author{W.M.~Lee}
\affiliation{Fermi National Accelerator Laboratory, Batavia, IL 60510}
\affiliation{Georgia State University, Atlanta, GA 30303}

\author{M.J.~Leitch}
\affiliation{Los Alamos National Laboratory, Los Alamos, NM 87545}

\author{N.~Makins}
\affiliation{Physics Division, Argonne National Laboratory, Argonne, IL 60439}
\affiliation{University of Illinois at Urbana-Champaign, Urbana, IL 61801}

\author{P.L.~McGaughey}
\affiliation{Los Alamos National Laboratory, Los Alamos, NM 87545}

\author{J.M.~Moss}
\affiliation{Los Alamos National Laboratory, Los Alamos, NM 87545}

\author{B.A.~Mueller}
\affiliation{Physics Division, Argonne National Laboratory, Argonne, IL 60439}

\author{P.M.~Nord}
\affiliation{Valparaiso University, Valparaiso, IN 46383}

\author{V.~Papavassiliou}
\affiliation{New Mexico State University, Las Cruces, NM 88003}

\author{B.K.~Park}
\affiliation{Los Alamos National Laboratory, Los Alamos, NM 87545}

\author{G.~Petitt}
\affiliation{Georgia State University, Atlanta, GA 30303}

\author{M.E.~Sadler}
\affiliation{Abilene Christian University, Abilene, TX 79699}

\author{W.E.~Sondheim}
\affiliation{Los Alamos National Laboratory, Los Alamos, NM 87545}

\author{P.W.~Stankus}
\affiliation{Oak Ridge National Laboratory, Oak Ridge, TN 37831}

\author{T.N.~Thompson}
\affiliation{Los Alamos National Laboratory, Los Alamos, NM 87545}

\author{R.S.~Towell}
\affiliation{Abilene Christian University, Abilene, TX 79699}

\author{R.E.~Tribble}
\affiliation{Texas A\&M University, College Station, TX 77843}

\author{M.A.~Vasiliev}
\affiliation{Texas A\&M University, College Station, TX 77843}

\author{J.C.~Webb}
\affiliation{New Mexico State University, Las Cruces, NM 88003}

\author{J.L.~Willis}
\affiliation{Abilene Christian University, Abilene, TX 79699}

\author{D.K.~Wise}
\affiliation{Abilene Christian University, Abilene, TX 79699}

\author{G.R.~Young}
\affiliation{Oak Ridge National Laboratory, Oak Ridge, TN 37831}

\collaboration{FNAL E866/NuSea Collaboration}
\noaffiliation

\date{\today}

\begin{abstract}
We report a measurement of the angular distributions of Drell-Yan
dimuons produced using an 800 GeV/c proton beam on a hydrogen 
target. The polar and azimuthal angular distribution parameters
have been extracted over the kinematic 
range $4.5 < m_{\mu \mu} 
< 15$ GeV/c$^2$ (excluding the $\Upsilon$ resonance region),
$0 < p_T < 4 $ GeV/c, and $0 < x_F < 0.8$. The $p+p$ angular distributions
are similar to those of $p+d$, and both data sets are compared
with models which attribute the $\cos 2 \phi$ distribution 
either to the presence of the transverse-momentum-dependent Boer-Mulders 
structure function $h_1^\perp$ or to QCD effects. The data indicate
the need to include QCD effects before reliable information on
the Boer-Mulders function can be extracted. The validity of 
the Lam-Tung relation 
in $p+p$ Drell-Yan is also tested. 

\end{abstract} 

\maketitle
The study of the transverse momentum dependent (TMD) parton 
distribution functions of the nucleon has received much attention 
in recent years as it provides new perspectives on the hadron
structure and QCD~\cite{barone02}. One of these TMD distribution
functions, first 
considered by Sivers~\cite{sivers90}, represents 
the correlation between the quark's transverse momentum, $k_\perp$, 
and the transverse spin of the nucleon, $S_\perp$. This so-called 
Sivers function, $f^\perp_{1T}(x,k^2_\perp)$, where $x$ is the
fraction of proton's momentum carried by the quark, is time-reversal odd
(T-odd) and can arise from initial- or 
final-state interactions~\cite{brodsky02}.
More generally, the requirement of gauge
invariance of parton distributions was shown to provide nontrivial 
phases leading to the existence of T-odd distribution 
functions~\cite{collins02,ji02}. Recent measurements of the semi-inclusive
deep-inelastic scattering (SIDIS) by the HERMES~\cite{hermes05} 
and COMPASS~\cite{compass05} collaborations
have shown clear evidence for the presence of the T-odd Sivers functions.
These data also allow the first determination~\cite{vogelsang05} of 
the magnitude and flavor structure of the Sivers functions.

Another T-odd distribution function is the Boer-Mulders function, 
$h^\perp_1(x,k^2_\perp)$, which signifies the correlation 
between $k_\perp$ and the quark transverse spin, $s_\perp$, in an unpolarized 
nucleon~\cite{boer98}. The Boer-Mulders function is the 
chiral-odd analog of the Sivers function and
also owes its existence to the presence of initial/final state 
interactions~\cite{boer03}. While the Sivers function is 
beginning to be quantitatively determined from
the SIDIS experiments, very little is known about the Boer-Mulders
function so far.

Several model calculations have been carried out for the 
Boer-Mulders functions. In the quark-diquark model, it was
shown that the Boer-Mulders functions are identical to
the Sivers functions when only the scalar diquark configuration 
is considered~\cite{boer03, gamberg03}. More recently, calculations 
taking into account both
the scalar and the axial-vector diquark configurations found significant
differences in flavor dependence between the Sivers and Boer-Mulders 
functions~\cite{gamberg08}.
In particular, the $u$ and $d$ valence quark Boer-Mulders 
functions are predicted
to be both negative, while the Sivers function is negative for the $u$ 
and positive for the $d$ valence quarks. Other calculations 
using the MIT bag model~\cite{yuan03}, the relativistic constituent 
quark model~\cite{pasquini07}, the large-$N_c$ model~\cite{pobylitsa03}, 
and lattice QCD~\cite{gockeler07} 
also predict negative signs for the $u$ and $d$ valence Boer-Mulders
functions. Burkardt recently pointed out~\cite{burkardt08} 
that the negative signs for the Boer-Mulders functions are
expected for both nucleons and pions. The model predictions
for the same signs of the $u$ and $d$ Boer-Mulders functions 
remain to be tested experimentally. Furthermore, the striking 
prediction~\cite{collins02} that the T-odd Boer-Mulders functions
in the SIDIS process will change their signs for the Drell-Yan
process also awaits experimental confirmation.

The Boer-Mulders functions can be extracted~\cite{boer99}
from the azimuthal angular distributions in the unpolarized Drell-Yan
process, $h_1 h_2 \to l^+ l^- x$. The general expression for the Drell-Yan
angular distribution is~\cite{lam78}
\begin{equation}
\frac {d\sigma} {d\Omega} \propto 1+\lambda \cos^2\theta +\mu \sin2\theta
\cos \phi + \frac {\nu}{2} \sin^2\theta \cos 2\phi,
\label{eq:eq1}
\end{equation}
\noindent where $\theta$ and $\phi$ are the polar and azimuthal decay angle
of the $l^+$ in the dilepton rest frame. Boer showed that the $\cos 2\phi$
term is proportional to the convolution of the quark and antiquark
Boer-Mulders functions in the projectile and target~\cite{boer99}.
This can be understood by noting that the Drell-Yan cross
section depends on the transverse spins of the annihilating quark and 
antiquark. Therefore, a correlation between the transverse spin and
the transverse momentum of the quark, as represented by the Boer-Mulders
function, would lead to a 
preferred transverse momentum direction. 

Pronounced $\cos 2 \phi$ dependences
were indeed observed in the NA10~\cite{falciano86} and E615~\cite{conway89} 
pion-induced Drell-Yan experiments, and attributed to the 
Boer-Mulders function.
The first measurement of the $\cos 2 \phi$
dependence of the proton-induced Drell-Yan process was recently reported for 
$p+d$ interactions
at 800 GeV/c~\cite{zhu07}. In contrast to pion-induced Drell-Yan,
significantly smaller (but non-zero) cos$2\phi$ azimuthal angular dependence 
was observed in the $p+d$ reaction. While the pion-induced Drell-Yan process
is dominated by annihilation between a valence antiquark in the pion
and a valence quark in the nucleon, the
proton-induced Drell-Yan process involves a valence quark in the proton 
annihilating with a sea antiquark in the nucleon. Therefore, the
$p+d$ result suggests~\cite{zhu07} that the Boer-Mulders functions for 
sea antiquarks are significantly smaller than those for valence quarks. 

A recent analysis~\cite{zhang08} indicated that the E866 $p+d$
data are consistent with the $u$ and $d$ Boer-Mulders
functions having the same signs, as predicted by various models. However, the 
$p+d$ data alone cannot provide an unambiguous determination of the 
flavor dependence of the Boer-Mulders functions. Moreover, it was 
recently pointed
out~\cite{boer06,berger07} that QCD processes
would lead 
to a sizeable $\cos 2\phi$
effect which has not been taken into account in the 
extractions~\cite{boer99,zhang08,zhang08a} of Boer-Mulders
functions from the Drell-Yan data. In
this paper we report the Drell-Yan angular distributions of the
$p+p$ reaction at 800 GeV/c, which provides further constraints on
the flavor dependence of the Boer-Mulders functions~\cite{zhang08a}. 
We also compare the
$\cos 2\phi$ dependences of $p+p$ and $p+d$ data with the prediction of 
QCD.
\begin{figure}[tb]
\includegraphics*[width=\linewidth]{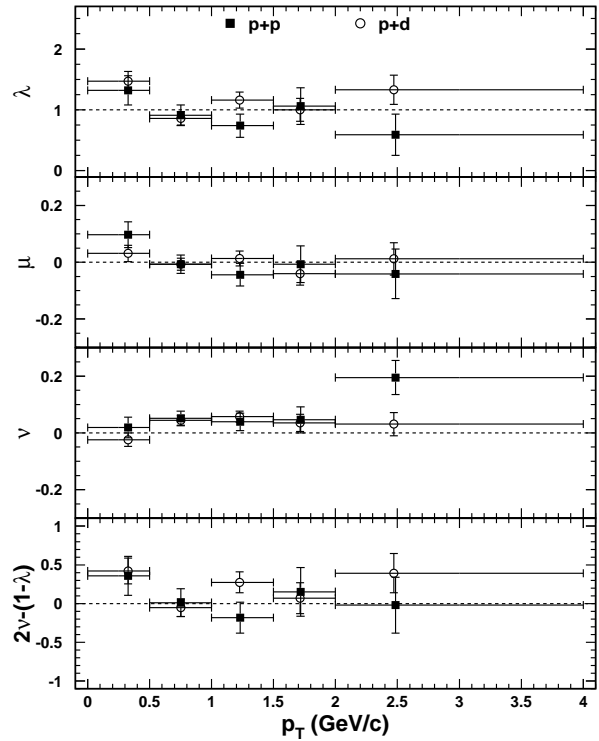}
\caption{ Parameters $\lambda, \mu, \nu$ and $2\nu - (1-\lambda)$
vs.\ $p_T$ in the Collins-Soper frame. Solid squares (open circles) 
are for E866 $p+p$ ($p+d$) at 800 GeV/c. The vertical error bars 
include the statistical uncertainties only.}
\label{ppfig1}
\end{figure}

The Fermilab E866 experiment was performed using the upgraded Meson-East 
magnetic pair spectrometer~\cite{hawker}. An 800 GeV/c primary proton 
beam with up to $2 \times 10^{12}$ protons per 20 s beam spill 
was incident upon one of three identical 50.8 cm long target flasks 
containing either liquid hydrogen, 
liquid deuterium or vacuum. A copper beam dump located inside the 
second dipole
magnet (SM12) absorbed protons that passed through the target. Downstream
of the beam dump was an absorber wall that 
removed hadrons produced in the target and the beam dump.

\begin{table}[tbp]
\caption {Mean values of the $\lambda, \mu, \nu$ parameters and the quantity
$2\nu -(1-\lambda)$ for the $p+p$, $p+d$, and $\pi^- +W$ 
Drell-Yan measurements.}
\begin{center}
\begin{tabular}{|c|c|c|c|}
\hline
\hline
 & $p+p$ & $p+d$ & $\pi^- +W$ \\
 & 800 GeV/c & 800 GeV/c & 194 GeV/c \\
 & (E866) & (E866) & (NA10) \\
\hline
$\langle \lambda \rangle$ & $0.85 \pm 0.10$ & $1.07 \pm 0.07$ & 
$0.83 \pm 0.04$ \\
\hline
$\langle \mu \rangle$ & $-0.026 \pm 0.019$ & $0.003 \pm 0.013$ & 
$0.008 \pm 0.010$ \\
\hline
$\langle \nu \rangle$ & $0.040 \pm 0.015$ & $0.027 \pm 0.010$ & 
$0.091 \pm 0.009$ \\
\hline
$\langle 2 \nu - (1-\lambda) \rangle$ & $-0.07 \pm 0.10$ & 
$0.12 \pm 0.07$ & $0.01 \pm 0.04$ \\
\hline
\hline
\end{tabular}
\end{center}
\end{table}

Several settings of the
currents in the three dipole magnets (SM0, SM12, SM3)
were used in order to optimize acceptance for different dimuon mass regions.
Data collected with the ``low mass" and ``high mass" 
settings~\cite{hawker} on
liquid hydrogen and empty targets were used in this analysis.
The detector system consisted of
four tracking stations and a momentum analyzing magnet (SM3). 
Tracks reconstructed by the drift chambers were extrapolated to the target
using the momentum determined from the bend angle in SM3.
The target position was used to refine the parameters of each muon track. 

From the momenta of the $\mu^+$ and $\mu^-$, kinematic variables of
the dimuons ($x_F, m_{\mu\mu}$, and $p_T$, where $x_F$ is the fraction of the
c.m. momentum carried by dimuon of mass $m_{\mu\mu}$, and $p_T$
is the dimuon transverse momentum) were readily reconstructed.
The muon angles $\theta$ and $\phi$ in the Collins-Soper
frame~\cite{collins77} were also calculated. To eliminate the $J/\Psi$
and $\Upsilon$ resonance background, dimuon events with $m_{\mu\mu} < 4.5$
GeV/c$^2$ and 9.0 GeV/c$^2 < m_{\mu\mu} < 10.7$ GeV/c$^2$ were rejected in
the analysis.  
A total of $\approx$54,000
$p+p$ Drell-Yan events covering the decay angular range $-0.5 < \cos\theta
<0.5$ and $-\pi < \phi < \pi$ remain. Detailed Monte Carlo simulations
of the experiment using the MRST98 parton 
distribution functions~\cite{mrst} for
NLO Drell-Yan cross sections have shown good agreement with the data for 
a variety of measured quantities.

Figure 1 shows the angular distribution parameters $\lambda, \mu,$ and
$\nu$ vs.\ $p_T$. To extract these parameters, the Drell-Yan data were
grouped into 5 bins in $\cos\theta$ and 8 bins in $\phi$ for each $p_T$
bin. A least-squares fit
to the data using Eq.~1 to describe the angular distribution was
performed. The extracted values of $\lambda, \mu, \nu$ are
insensitive to their values used in the Monte Carlo simulation.
Only statistical errors are shown
in Fig.~1. The primary contributions to the systematic errors are the 
uncertainties of the incident beam angles on target. Analysis 
performed by 
allowing the beam angles to vary within their ranges of uncertainty
has shown that the
systematic errors are small compared to the statistical errors. 
The E866 $p+d$ Drell-Yan data are also shown in Fig.~1 for comparison
with the E866 $p+p$ data. The $p+d$ data contain a total of
$\approx$118,000 events covering an identical $\cos\theta$ range.
The $p_T$-averaged values of 
$\langle\lambda\rangle, \langle\mu\rangle,$ and 
$\langle\nu\rangle$ for $p+p$, $p+d$, and the 
NA10 $\pi^- + W$ data~\cite{falciano86} 
are listed in Table~I.
Within statistics, the angular distributions of $p+p$ are consistent
with those of $p+d$. Also shown in Fig.~1 and Table~I is the quantity 
$2\nu -(1-\lambda)$, which should vanish if the Lam-Tung relation is
valid. While QCD effects
can lead to $\lambda \ne 1$ and $\mu, \nu \ne 0$, Lam and Tung
showed~\cite{lam80} that the relation $1-\lambda = 2\nu$ is largely
unaffected by QCD corrections. 
Table~I shows that while $\langle\lambda\rangle$ deviates from 1 
and $\langle\nu\rangle$ is nonzero for
the E866 $p+p$ and the NA10 $\pi^- + W$ Drell-Yan data, the Lam-Tung relation
is indeed quite well satisfied within statistical uncertainty for all
$p_T$. This differs 
from the observation of
a significant violation of the Lam-Tung relation at large $p_T$ 
by the E615 collaboration
in the $\pi^- + W$ reaction at 252 GeV/c~\cite{conway89}.

Figure 2 shows the parameter $\nu$ vs.\ $p_T$ for the $p+p$ and
$p+d$ Drell-Yan data. The solid curves are
calculations~\cite{zhang08,zhang08a} for $p+p$ and $p+d$ using
parametrizations of the Boer-Mulders functions deduced from a fit 
to the $p+d$ Drell-Yan data.
The predicted larger values of $\nu$ for $p+p$ compared 
to $p+d$ in the region of $p_T \sim 1.5$ GeV/c are not 
observed (the predicted $p+p/p+d$ ratio, $R$, for
$0.5 < p_T < 2.0$ GeV/c, is $\sim 2$, while the data give
$R=1.0 \pm 0.5$). Furthermore, the shape of
the predicted $p_T$ dependence differs
from that of the data, resulting in a reduced $\chi^2$
value of 3.2 for 5 degrees of freedom (probability of 0.7\%). 
This strongly suggests that there could
be other mechanisms contributing to the $\cos 2\phi$ azimuthal angular
dependence at large $p_T$. In recent papers~\cite{boer06,berger07},
the QCD contribution
to the $\cos 2\phi$ azimuthal angular dependence is given as
\begin{equation}
\nu = \frac {Q^2_\perp/Q^2} {1+\frac{3}{2}Q^2_\perp/Q^2},
\label{eq:eq2}
\end{equation}
\noindent where $Q_\perp$ is the dimuon transverse momentum. The
predicted QCD contribution, the same for $p+p$ and $p+d$ due to the
identical kinematic coverage for the two reactions, 
is shown as the dot-dashed curve in Fig.~2. A comparison
between the QCD
prediction with the data gives
a reduced $\chi^2$ of 1.0 for 5 degrees of freedom (probability
of 42\%) for $p+p$ and a reduced $\chi^2$ of 1.9 (probability of
9\%) for $p+d$. From Fig.~2 it is
evident that the QCD contribution is expected to become
more important at high $p_T$ while the
Boer-Mulders functions contribute primarily at lower $p_T$.
An analysis combining both effects is 
required in order to extract reliably the
Boer-Mulders functions from the $p+p$ and $p+d$ data.
It is worth noting that the $\pi^- +W$ Drell-Yan 
data~\cite{falciano86,conway89}
also show large values of $\nu$ at large $p_T$, consistent with 
the presence of QCD effects.

\begin{figure}[tb]
\includegraphics*[width=\linewidth]{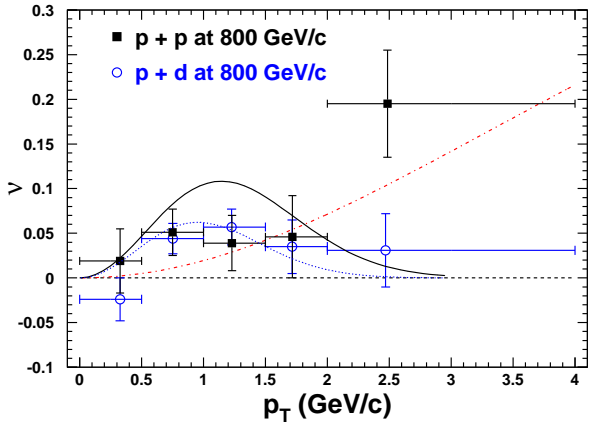}
\caption{(color online). Parameter $\nu$ vs.\ $p_T$ in the Collins-Soper
frame for the $p+p$ and $p+d$ Drell-Yan data. 
The solid and dotted curves are
calculations~\cite{zhang08} for $p+p$ and $p+d$, respectively, 
using parametrizations
based on a fit to the $p+d$ data. The dot-dashed curve is the
contribution from the QCD process (Eq. 2).}
\label{ppfig2}
\end{figure}

The $p+p$ Drell-Yan angular distributions have also been analyzed for other
kinematic variables. Figure 3 shows the values 
of $\nu$ vs.~$m_{\mu\mu}, x_F, x_1,$ and $x_2$, where $x_1$ 
and $x_2$ are the
Bjorken-$x$ for the beam and target partons, respectively. Again, for
each bin the data were divided into 5 bins in $\cos\theta$ and 8 bins in
$\phi$ in order to extract the angular distribution parameters. 
The $p+d$ data are also shown for comparison. Figure 3 shows
that the magnitude of $\nu$ for $p+p$ is consistent with that for $p+d$ for
most of the kinematic regimes. These data provide further input for
future extraction of the Boer-Mulders functions.

\begin{figure}[tb]
\includegraphics*[width=\linewidth]{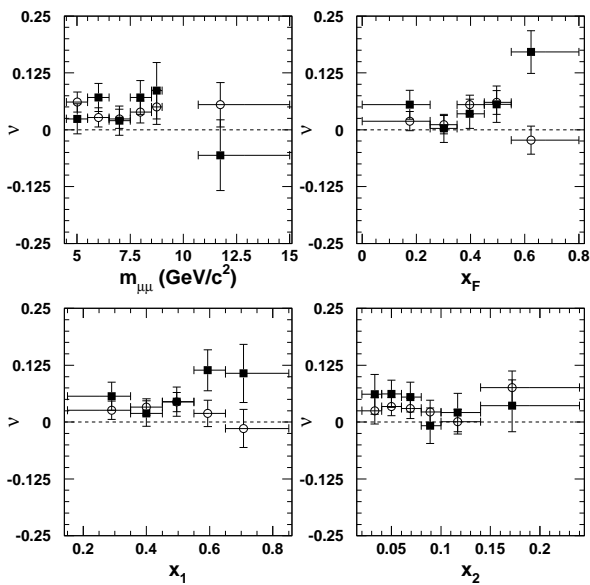}
\caption{Parameter $\nu$ vs.\ $m_{\mu\mu}$, $x_F$, $x_1$, 
and $x_2$ in the 
Collins-Soper frame for $p+p$ (solid squares) and 
$p+d$ (open circles) at 800 GeV/c. 
The vertical error bars correspond
to the statistical uncertainties only.}
\label{ppfig3}
\end{figure}

In summary, we report a measurement of the angular distributions of
Drell-Yan dimuons for $p+p$ at 800 GeV/c. The pronounced $\cos 2 \phi$
azimuthal angular dependence observed previously in pion-induced Drell-Yan
is not observed in the $p+p$ reaction. The Lam-Tung relation remains  
valid for the $p+p$ Drell-Yan data. The overall magnitude of the $\cos 2 \phi$
dependence for $p+p$ is consistent with, but slightly larger than 
that of $p+d$. The data suggest the presence
of higher-order QCD corrections at high $p_T$, and it is important to
take this contribution into account before reliable extraction of the 
Boer-Mulders functions could be obtained. 

We acknowledge helpful discussion with Bo-Qiang Ma, Bing Zhang, Matthias
Burkardt, Feng Yuan, Werner Vogelsang, and Jianwei Qiu.
This work was supported in part by the U.S. Department of Energy and 
the National Science Foundation.

\end{document}